# Electron-beam ignited, high-frequency driven vacuum ultraviolet excimer light source

[1]T. Dandl, [2]H. Hagn, [1]T. Heindl, [1]R. Krücken, [3]J. Wieser, and [1]A. Ulrich

[1]Physik Department E12, Technische Universität München, James-Franck-Str. 1, 85748 Garching, Germany
[2]Physik Department E15, Technische Universität München, James-Franck-Str. 1, 85748 Garching, Germany
[3]excitech GmbH, Branterei 33, 26419 Schortens, Germany

**Abstract.** Transformation of a table-top electron beam sustained 2.45 GHz RF discharge in rare gases into a self burning discharge has been observed for increasing RF-amplitude. Thereby, the emission spectrum undergoes significant changes in a wide spectral range from the vacuum ultraviolet (VUV) to the near infrared. A strong increase of VUV excimer emission is observed for the self burning discharge. The so called first excimer continuum, in particular, shows a drastic increase in intensity. For argon this effect results in a brilliant light source emitting near the 105 nm short wavelength cutoff of LiF windows. The appearance of a broad-band continuum in the UV and visible range as well as effects of RF excitation on the atomic line radiation and the so called third excimer continuum are briefly described.

Brilliant vacuum ultraviolet (VUV) light sources have been developed for example for photo ionization of analyte molecules in mass spectrometry. The "soft" photo ionization has the advantage to minimize fragmentation of the molecules allowing also complex mixtures of substances to be analyzed [1,2]. Excimer light sources in which a low energy electron beam is used to excite the rare gases argon and krypton [3,4] have been found to be particularly well suited for this application. They emit predominantly the so called second excimer continuum which results from the radiative decay of excimer molecules in their lowest vibrational level. A comparison of electron beam induced emission spectra of helium, neon, argon, krypton, and xenon at 1 bar gas pressure is shown in Fig. 1. Helium and neon can not be used for VUV-light applications in vacuum since there are no output windows available below the wavelength-cutoff of LiF or MgF$_2$ around 105 and 110 nm, respectively. As can be seen in Fig. 1 there is a gap in the rare gas excimer spectra between 105 nm and the onset of the argon emission at about 120 nm for light sources emitting only the second excimer continuum. It is however important to cover the related photon energy range from about 10 to 12 eV because the ionization energy of many important analyte molecules lies in this energy range. Security relevant substances for example such as explosives fall into this category [5].

Here we present a way to extend the emission of excimer light sources into this wavelength- and photon-energy range. The concept is to increase the intensity of the so called first excimer continuum of argon. An argon spectrum with an intense first continuum as described in this paper is shown in the argon-chart of Fig. 1. The first continuum is related to optical transitions between high lying vibrational levels of the excimer molecules and the repulsive ground state. It extends from the resonance lines of the rare gases towards the second continuum (see Fig. 2 for a level diagram). Therefore it starts near 106.7 nm (11.6 eV) for argon. It will be shown that it is possible to operate an RF discharge in such a mode that this continuum reaches the intensity level of the second continuum which dominates the VUV emission spectrum of rare gases for sole electron beam excitation as shown in Fig. 1.

Dense gases, typically at atmospheric pressure and above are required for efficient formation of excimer molecules [6] and electron beam excitation is a well suited excitation method. It is known from particle excitation of rare gases that the electron energy distribution function (EEDF) peaks at very low electron energy [7,8]. The concept to intensify the first excimer continuum is to modify the EEDF in such a way that these low energy electrons are accelerated by an RF field. With this enhanced energy they can redistribute the population of the vibrational levels of the excimer molecules via collisions. Gas heating accompanied by this heating of the electron component will act in the same direction.

The experimental setup is shown schematically in Fig. 3 and will be discussed in more detail in a forthcoming publication. Briefly, an electron beam of 12 keV particle energy was formed with a small electron gun and sent through a thin ceramic membrane (silicon nitride and silicon oxide, 300 nm thick) from vacuum into a rare gas target at about 1 bar gas pressure. The gas was continuously purified using a metal bellows compressor for gas circulation and a rare gas purifier (SAES Monotorr) to obtain clean excimer spectra as shown in Fig 1. A flat aluminum electrode was mounted at variable distance in front of the membrane and connected to a 2.45 GHz RF generator. The RF power delivered by the generator ranged from 0 to 30 W. The generator was protected from RF power reflected at the target cell by a circulator. RF power coupled into the gas was determined with a directional coupler by measuring the difference of RF-power flowing towards the target cell and the reflected RF power, respectively.

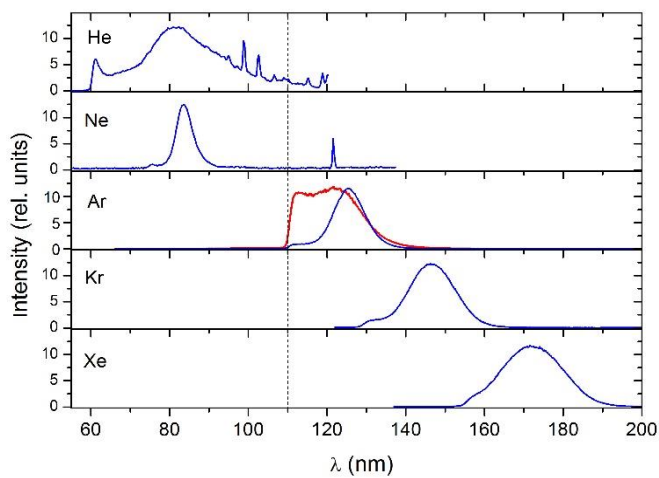

**Fig. 1**. *Emission spectra of the rare gases in the vacuum ultraviolet spectral region are shown. The spectra were recorded using 10 keV electron beam excitation of gas targets at 1 bar gas pressure. The so called second excimer spectra dominate the emission with conversion efficiencies of electron beam power into light up to ~40 %. The first continuum is visible as a short wavelength "shoulder" of the second continuum. The novel operation of the source described in this paper is shown for argon (red spectrum, scaled in intensity to e-beam spectrum). The cut-off wavelength of MgF$_2$ windows at 110 nm is marked by a dashed line.*

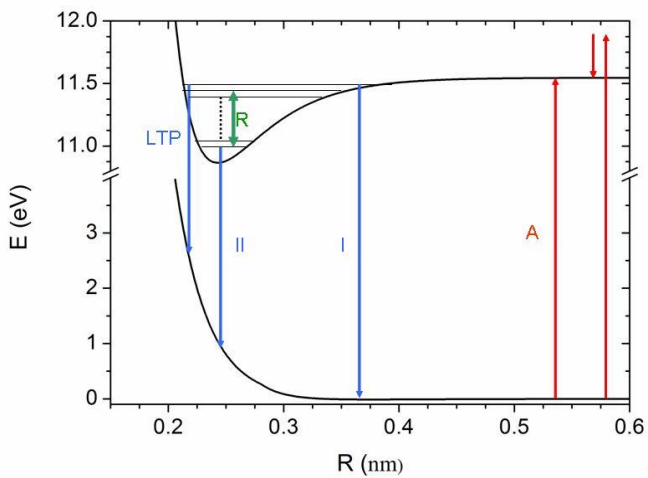

**Fig. 2.** *Schematic level diagram of the lowest lying atomic and molecular states of argon. Transitions from vibrationally relaxed molecules to the repulsive ground state are marked II. These transitions lead to the so called second excimer continuum of the rare gases. Transitions marked I are related to the so called first continuum and the "classical left turning point, LTP" (left side). Direct excitation (A), ionization, and recombination is schematically shown on the right side of the diagram. A way to enhance the intensity of the first continuum by a collisional redistribution of the population of vibrational levels (R) is described in this paper.*

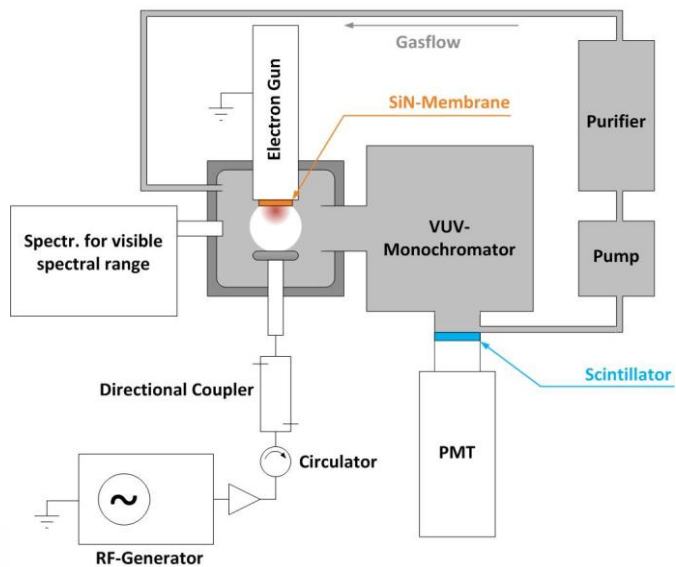

**Fig. 3.** *A schematic drawing of the experimental setup is shown. The target cell was built on the basis of a 40 mm diameter vacuum double cross (CF 40).*

Light emission from the target cell was observed with a windowless VUV/XUV scanning monochromator (Acton, Model VM-502) which was implemented as an integral part of the target cell gas system (see Fig. 3). A tetraphenylbutadien scintillator and a phototube operated in the counting mode was used as detector.

Spectra emitted from 1.3 bar argon in the vacuum ultraviolet wavelength range are shown in Fig. 4 for three different conditions. Spectrum A was recorded with 5 µA (0.05 W) electron beam excitation, alone. Spectrum B shows the modification when about 0.5 W RF power is coupled into the gas in addition to the electron beam power. For this RF power level and the electrode geometry described above, the light emission stops when the electron beam is switched off (beam sustained discharge). Spectrum C, finally, shows a self burning RF discharge driven by about 6 W RF power at 2.45 GHz coupled in to the gas. This input power level was a limit for the present setup due to RF losses in the cable and an impedance mismatch at the target cell. Important observations were that the volume remains rather restricted and that the VUV excimer light output exceeds the VUV emission with sole electron beam excitation by a factor of ~10. Note that the overall VUV light intensity is first reduced when RF power is added to the electron beam excited gas, as in the case of a beam sustained discharge.

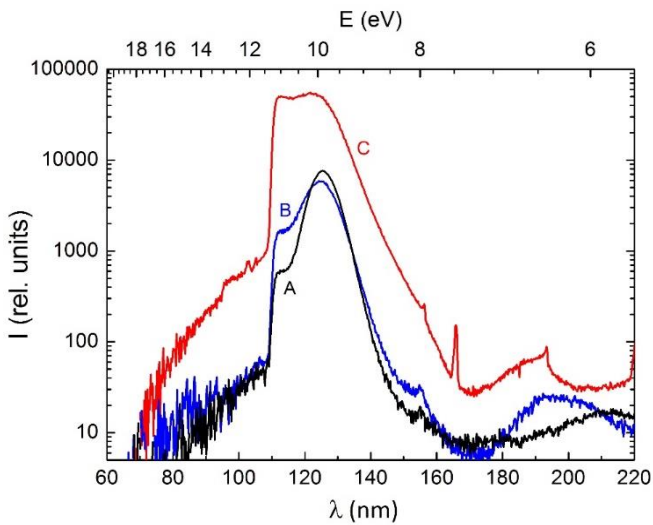

**Fig. 4.** *VUV - emission spectra of argon at 1.3 bar gas pressure are shown for three different conditions: Trace A shows a spectrum recorded with gas excitation by a 10 keV, 5 µA electron beam. Trace B shows a spectrum recorded with excitation by the same electron beam as trace A but with ~0.5 W (2.45 GHz) RF additional power coupled into the gas. Trace C shows a self burning discharge in the same gas excited with 6 W RF power and with the electron beam still in operation.*

The purely electron beam excited spectrum A in Fig 4 represents an emission where almost all excimer molecules are relaxed to the lowest vibrational level. Spectrum B in Fig. 4 is interpreted by a redistribution in the population density of the excimer molecules with respect to the vibrational levels, as discussed above. Population of the higher lying levels leads to enhanced emission in the first continuum and the so called "classical left turning point" of the molecular oscillation in high vibrational levels (see Fig. 2). The overall light emission is reduced by about 25 % with increasing RF power until the self burning discharge starts and this trend is reversed. From that point on the VUV emission rises linearly with the applied RF power and the trend that the first continuum emission increases in intensity with respect to the second continuum intensity prevails. This effect could be observed up to the maximum RF power of 6 W coupled into the gas. Scaling our results of an efficiency measurement for pure electron beam excitation of excimer light [9] we conclude that the argon source emitted up to 400 mW in the VUV region, integrating over first and second continuum. This corresponds to an efficiency of about 6 %. Subtracting a scaled second continuum spectrum from spectrum C in Fig. 4 we conclude that up to 180 mW were emitted in the short wavelength first continuum region. In this region an intensity increase by more than a factor of 100 was observed in comparison with spectrum B in Fig. 4.

The effect of RF power in the ultraviolet, visible, and near infrared spectral range is demonstrated in Fig. 5. Data were recorded using a compact grating spectrometer with fiber optics input (Ocean Optics QE65000). A broad-band continuum is observed whenever RF power is coupled into the gas. Mainly from time resolved measurements performed with a pulsed, combined electron beam – RF discharge we attribute the origin of this continuum to photo-recombination [10]. This and the following observations will be described in detail in a forthcoming publication. The atomic line radiation is enhanced by the RF power. Analog effects were observed also with krypton as the light emitting medium. The so called third excimer continuum [11] can only be observed in a combined mode of excitation when the electron beam is on as in Fig. 4. It disappears completely when the electron beam is switched off.

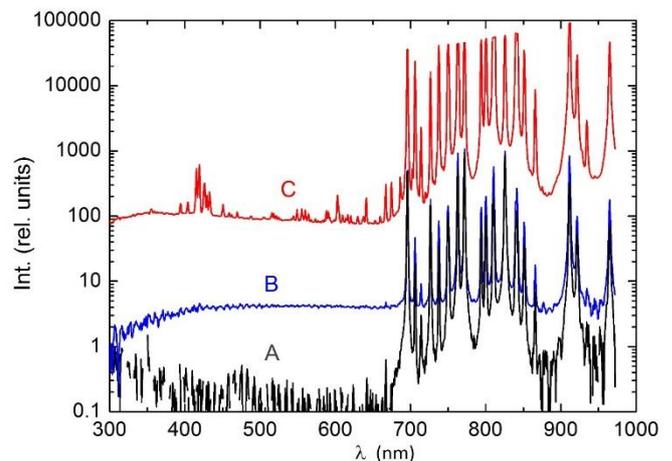

**Fig. 5.** *Emission spectra of argon in the ultraviolet and visible spectral range are shown. Trace A refers to 0.05 W electron beam excitation, trace B to an electron beam sustained discharge with 0.05 W electron beam power and ~0.5 W RF power, and trace C to a self burning 6 W RF discharge. A broad-band continuum emission appears in addition to the atomic line radiation when RF power (2.45 GHz) is applied (see text).*

For application in brilliant light sources it is important to know the geometry of the light emitting volume. The photographs shown in Fig. 6 give an impression of this parameter. So far no quantitative study of the geometrical effects and their wavelength dependence has been performed. The visible light emitting volume of the self–burning discharge is about 20 mm$^3$ for 7 mm electrode distance at the highest RF power levels (6 W) used here (Fig. 6 C). Two additional effects can be observed visually: The color of the emission changes from dark violet (in the case of argon) to a whitish glow (Fig. 6) and the visible light is strong near the electrodes as it is typical for many discharges ("cathode- and anode fall", Fig. 6 C). We assume that the VUV emission is mainly emitted from the more homogeneously glowing middle section of the discharge but so far we can not exclude that the first excimer continuum comes predominantly from the probably hotter regions near the electrodes. It is interesting to note that the high frequency discharge causes no damage to the thin ceramic entrance foil for the electron beam. This is similar to the advantages of RF discharges with respect to electrode sputtering as for example discussed in reference [12].

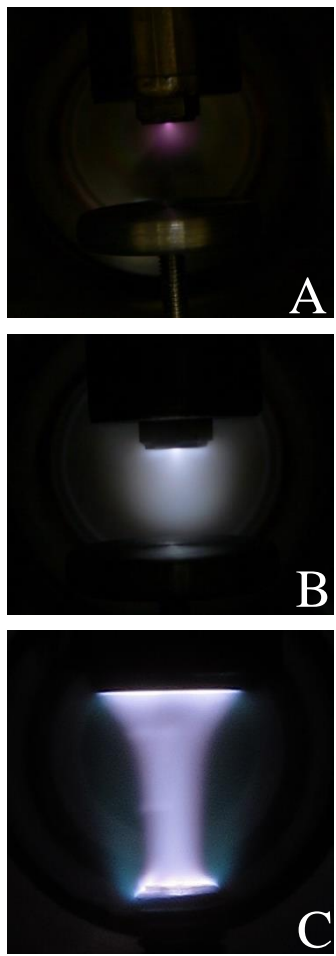

***Fig. 6.*** *The development of the visual appearance of the light emission of argon from electron beam excitation (A) via an electron beam sustained discharge (B) to a self burning discharge (C) is documented by the three photographs. These photographs correlate with the spectra A, B, and C in Fig. 5. In A and B the gas pressure in the cell was 0.5 bar, the electrode distance 10 mm, and its diameter also 10 mm. These values in C were 1 bar, 7 mm, and 3 mm, respectively.*

In summary, we have demonstrated first a novel, table top operation of an electron beam sustained discharge in rare gases. Second, the setup has been used for igniting a self burning RF discharge. It was shown that the excimer emission in the vacuum ultraviolet wavelength region is the dominant feature in the spectra in both cases and that the emission of the first continuum is strongly enhanced with increasing RF power. The self burning RF discharge in argon represents, thereby, to our best knowledge the most brilliant table top VUV source close to the short wavelength cut-off of LiF and $MgF_2$ windows used in VUV optics.

## Acknowledgement

This work has been funded by the German Federal Ministry of Research (BMBF) und contract No. 13N9528.